\documentstyle[epsfig]{aipproc}

\def\be{\begin{equation}}

\def\ee{\end{equation}}

\def\bea{\begin{eqnarray}}

\def\eea{\end{eqnarray}}

\newcommand{\ba}{\begin{array}}
\newcommand{\ea}{\end{array}}

\def\lsim{\mathrel{\rlap{\lower4pt\hbox{\hskip1pt$\sim$}}
    \raise1pt\hbox{$<$}}}         
\def\gsim{\mathrel{\rlap{\lower4pt\hbox{\hskip1pt$\sim$}}
    \raise1pt\hbox{$>$}}}         

\def\Pom{{\bf I\!P}}

\def\lsim{\mathrel{\rlap{\lower4pt\hbox{\hskip1pt$\sim$}}
    \raise1pt\hbox{$<$}}}         
\def\gsim{\mathrel{\rlap{\lower4pt\hbox{\hskip1pt$\sim$}}
    \raise1pt\hbox{$>$}}}         

\def\Pom{{\bf I\!P}}

\begin{document}

\title{Phenomenology of Diffractive DIS}

\author{\underline{Nikolai N. Nikolaev}$^{a),b),c)}$ 
and Bronislav G.Zakharov$^{c)}$}

\address{$^{a)}$ITKP der Universit\"at Bonn, Nusallee 14-16, D-53115 Bonn\\
$^{b)}$IKP, KFA J\"ulich, D-52425 J\"ulich, Germany\\
$^{c)}$L.D.Landau Institute, Kosygina 2, 1117 334 Moscow, Russia}
\maketitle




\begin{abstract}
The recent progress in the QCD theory of diffractive DIS (DDIS) is
reviewed. We place emphasis on pQCD scales, diffractive factorization
breaking, jet and charm production and new fundamental observables 
which are becoming accessible with the 
Leading Proton Spectrometers (LPS) of ZEUS and H1.
\end{abstract}

\section*{Diffractive structure functions}

This is an opening talk on diffraction at DIS'97 and we start 
with definitions.
The longitudinal (L) and the two transverse (T) polarizations 
of the exchanged
photon define the four components $d\sigma_{i}^{(3)}
(\gamma^{*}p\rightarrow p'X)/dM^{2}dp_{\perp}^{2}d\phi$ 
($i=T,L,TT'$ and $LT$) in the expansion
\bea
Q^{2}x{d\sigma^{(5)}(ep\rightarrow e'p'X)\over 
dQ^{2}dxdM^{2}dp_{\perp}^{2}d\phi} = {\alpha_{em} \over \pi}
\left\{(1-y+{1\over 2}y^{2})\cdot d\sigma_{T}^{D(3)}+
(1-y)\cdot d\sigma_{L}^{(3)}\right. \nonumber\\ 
\left. +
(1-y)\cos 2\phi \cdot d\sigma_{TT'}^{D(3)}+(2-y)\sqrt{1-y}\cos\phi 
\cdot d\sigma_{LT}^{D(3)}\right\}/dM^{2}dp_{\perp}^{2}d\phi
\label{eq:1.1}
\eea
where $M$ is the diffractive mass, $\vec{p}_{\perp}$
is the $(p,p')$ momentum transfer, and $\phi$ is the
angle between the $(e,e')$ and $(p,p')$ planes. In the charged current
(CC) 
DDIS there are also the  C- and P-odd,
 $TT'$ and 
$LT'$, terms \cite{BGNZcc}. 
Each $d\sigma_{i}^{(3)}$ 
defines a set of 
{\bf dimensionless} diffractive structure functions $F_{i}^{D(5)}$,
\be
{2\pi(Q^{2}+M^{2})d\sigma_{i}^{(3)}(\gamma^{*}p\rightarrow p'X)\over 
dM^{2}dp_{\perp}^{2}d\phi}={\sigma_{tot}^{pp}\over 16\pi}\cdot
{4\pi^2 \alpha_{em} \over Q^{2}}\cdot F_{i}^{D(5)}(\phi,p_{\perp}^{2},
x_{\Pom},\beta,Q^{2})\, ,
\label{eq:1.2}
\ee
$F_{i}^{D(4)}(p_{\perp}^{2},x_{\Pom},\beta,Q^{2})=\int {d\phi\over
2\pi}F_{i}^{D(5)}$ and 
$
F_{i}^{D(3)}(x_{\Pom},\beta,Q^{2})= 
{\sigma_{tot}^{pp}\over 16\pi }\int dp_{\perp}^{2}
F_{i}^{D(4)}\, .
$
The so-defined $F_{i}^{D(3)}$ are smooth functions of 
$x_{\Pom},\beta,Q^{2}$, 
in contrast to the ZEUS/H1 definition of $F_{2}^{D(3)}$ \cite{ZEUSF2Pom} 
which
blows up at $x_{\Pom}\rightarrow 0$ due to the extra factor $1/x_{\Pom}$. 
Here $\beta =Q^{2}/(Q^{2}+M^{2})$ and $x_{\Pom}=
x/\beta$ are the DDIS variables.

The microscopic QCD mechanism of DDIS, as developed 
in 1991 by Nikolaev and Zakharov \cite{NZ91,NZ92,NZ94}, is a grazing, 
quasielastic scattering of multiparton Fock states of the $\gamma^{*}$ 
on the proton. DDIS is best described viewing $\gamma^{*}$ 
as a system of color dipoles spanning between quarks, antiquarks 
and gluons \cite{NZ91,NZ92,NZ94}, interaction of which with the 
target proton is described by the color dipole cross section 
$\sigma(x,r)={\pi^{2}\over 3}r^{2}\alpha_{S}(r)G(x,q^{2}\approx 10/r^{2})$
\cite{NZglue}.
Here $G(x,q^{2})$ is the gluon structure function of the target nucleon 
and one sums the Leading Log${1\over x}$ pQCD diagrams for the color singlet 
two-gluon exchange. Systematic expansion of 
DDIS in excitations of the $q\bar{q},
q\bar{q}g$ and higher Fock states of the photon has been
developed in
\cite{NZ92,NZ94,NZsplit,GNZ95,GNZA3Pom,GNZcharm,GNZlong}. The
technique \cite{NZ92,NZ94,NZsplit} has been a basis of all the 
subsequent pQCD works on DDIS \cite{Bartels}. The Buchmuller et al.
model \cite{Buchmuller} is essentially 
identical to the color dipole picture; their function $W(r)$  is
an exact counterpart of our $\sigma(x,r)$ evaluated
in the pQCD Born approximation.
Bjorken's aligned jet model \cite{Bjorken} is an integral part of  
the color dipole approach. For the 1996 status report at 
DIS'96 see \cite{NZDIS'96}.

This extended presentation reviews also the
new results reported at DIS'97. 


\section*{Dipole sizes and pQCD scales in DDIS}

Vector meson production $\gamma^{*}p\rightarrow Vp$ is an exclusive 
DDIS. For the shrinkage of $\gamma^{*}$  the typical dipole size 
(the scanning radius) contributing 
to the production amplitude equals $r_{S}=
6/\sqrt{Q^{2}+m_{V}^{2}}$ and one finds \cite{KNNZ93,KNNZ94}: 
\be
\sigma_{L} \propto Q^{2}r_{S}^{2}\sigma^{2}(x_{\Pom},r_{S})
\propto G^{2}(x_{\Pom},q_{L}^{2})Q^{2}/(Q^{2}+m_{V}^{2})^{4}
\, .
\label{eq:2.1}
\ee
For the factor 6 in $r_{S}$ the relevant pQCD scale,
$q_{T,L}^{2}=\tau_{T,L}(m_{V}^{2}+Q^{2})$,  is rather small:
$\tau_{L}=0.1$-0.2 and $\tau_{T}=0.07$-0.15 \cite{NNZvector}.
Eq. ~(\ref{eq:2.1}) is a basis of the very successful pQCD 
phenomenology of 
vector meson production (\cite{NNZvector,NNPZdipole,NNPZlight} 
and references
therein). The exponent $\Delta_{eff}$ in 
$\sigma_{L} \sim (W^{2})^{2\Delta_{eff}}$, rises with 
$Q^{2}$ due to the rise of the pQCD scales $q_{L,T}^{2}$, in
 good
agreement with the experiment.

DDIS into continuum at $\beta \gsim$ 0.1--0.2 
is dominated by the $q\bar{q}$ excitation.
If $\vec{k}$ is the transverse momentum of the
$q$ jet w.r.t. the photon, then  \cite{NZ92,NZsplit,GNZcharm,GNZlong} 
\bea
{dF_{T}^{D(3)} \over dk^{2}} \sim
\left(1-{2(k^{2}+m_{f}^{2} \over M^{2}}\right)\cdot
{1\over (k^{2}+m_{f}^{2})^{2}}\cdot
G^{2}(x_{\Pom},q_{T}^{2} = {m_{f}^{2}+k^{2} \over (1-\beta)})\, ,
\nonumber \\
{dF_{L}^{D(3)} \over dk^{2}} \sim {1\over Q^{2}}\cdot
{1\over (k^{2}+m_{f}^{2})}\cdot
G^{2}(x_{\Pom},q_{L}^{2} = {m_{f}^{2}+k^{2} \over (1-\beta)})\, .
\label{eq:2.2}
\eea
These results have been re-derived by several groups \cite{Bartels}.
Because $q_{T,L}^{2}\propto k^{2}$, we predict strong enhancement
of large-$k$ jets. Furthermore, $F_{L}^{D(3)}$ is dominated by hard
jets and 
scaling violations almost entirely compensate the higher twist
behavior of $F_{L}^{D(3)}\propto
{1\over Q^{2}}\beta^{3}(1-2\beta)^{2} G^{2}(x_{\Pom},{1\over 4\beta}Q^{2})$
\cite{GNZlong} which, for $\beta \gsim 0.9$, completely takes 
over the leading twist transverse structure function \cite{NZ92,GNZcharm}, 
$
F_{T}^{D(3)}\propto {1\over m_{f}^{2}}\beta(1-\beta)^{2}(3+4\beta+8\beta^{2}) 
G^{2}(x_{\Pom},q_{T}^{2}\approx {m_{f}^{2}\over (1-\beta)})$.
The applicability of pQCD improves for large $\beta$.
One often fits $F_{2}^{D(3)} \propto x_{\Pom}^{2(1-\alpha_{\Pom})}$.
The rise of our fixed-$M$ pQCD scale $q_{T}^{2}\sim {1\over 4}
m_{V}^{2}(1+{Q^{2}\over M^{2}})$ predicts the rise of $\alpha_{\Pom}$ 
with $Q^{2}$. Furthermore, at $Q^{2} \gsim 10 M^{2}$ one enters the 
$F_{L}^{D}$ dominated region of $q_{L}^{2} \sim {1\over 4}Q^{2}$ 
and $\alpha_{\Pom}$ must become as large as for vector meson production, 
which is indeed seen in the fixed-$M$ data from ZEUS \cite{Grothe}.
This is one of the manifestations of the duality between the continuum 
and vector meson production derived in 
\cite{GNZlong}.
For fixed $\beta$, the exponent $\alpha_{\Pom}$ must not depend on $Q^{2}$.

Production of $q\bar{q}$ dijets in DDIS at $\beta \ll 1$, and/or
by real photons, is especially interesting. In 1994 we showed that
in this case the transverse momentum 
$k$ of jets probes directly the transverse momentum
of gluons in the pomeron \cite{NZsplit} and $d\sigma_{T}^{D}
\propto |\partial G(x_{\Pom},k^{2})/\partial \log k^{2}|^2$ .

\begin{figure}[t!] 
\centerline{\epsfig{file=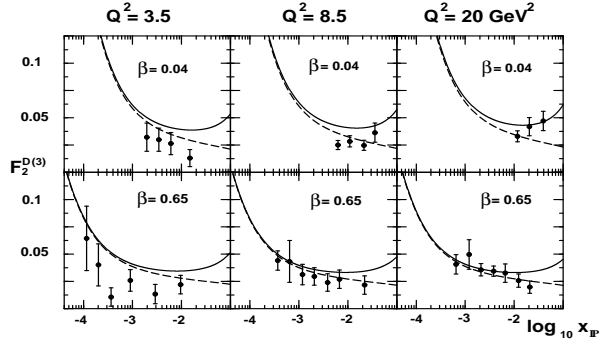,height=1.8in,width=3.1in}}
\vspace{10pt}
\caption{The dashed curve is the 1994 predictions for 
 $F_{2}^{D(3)}$ from the color dipole 
model {\protect\cite{GNZ95}}, the solid curve includes 
the $f$-reggeon and pion
exchange evaluated in {\protect\cite{NSZtriple}}, the experimental data 
are from  H1 {\protect\cite{Dirkmann}}.}
\label{fig1}
\end{figure}
  
DDIS at $\beta \lsim $ 0.1--0.2  comes from excitation of the
$q\bar{q}g$ and higher Fock states of the photon. It is DIS off the
pomeron sea with the typical sea structure function, $
F_{T}^{D(3)}(x_{\Pom},\beta,Q^{2})\sim  
G^{2}(x_{\Pom},
q_{sea}^{2}\approx \mu_{G}^{2})\,,$
which obeys an approximate triple-pomeron factorization conjectured in
\cite{NZ91,NZ92} and derived in \cite{NZ94,GNZA3Pom},
\be
F_{2}^{D(4)}(p_\perp^{2}=0,x_{\Pom},\beta,Q^{2})
\approx {16\pi A_{3\Pom} \over \sigma_{tot}^{pp}}
F_{2p}(x_{\Pom},q_{sea}^{2}) \, ,
\label{eq:2.4}
\ee
which is
similar to the  triple-pomeron limit of soft diffraction 
$(a=p,\pi,K,\gamma)$
$
\left.M^{2}d\sigma(ap\rightarrow p'X)/
dp_{\perp}^{2} dM^{2} \right|_{p_{\perp}^{2}=0} = 
A_{3\Pom} \sigma_{tot}^{ap}\cdot \, .
$
Here $R_{c}=1/\mu_{G}\sim$ 0.2-$0.3 fm$ is a
propagation radius for perturbative gluons. In \cite{GNZA3Pom}
we showed that although
$A_{3\Pom}\approx$ 0.15-0.2 GeV$^{-2}$ is the dimensional quantity, it 
does not change much from $Q^{2}=0$ to DIS; parametrically 
$A_{3\Pom}\propto R_{c}^{2}$. The normalization of
the pomeron sea structure function {\bf predicted} in \cite{NZ92,GNZ95}
is in  good agreement with the experiment, see \cite{ZEUSF2Pom}
and  Fig.~1.

CC DDIS has several peculiarities \cite{BGNZcc}:
First, both the $F_{T}^{D}(3)$ 
and  $F_{L}^{D}(3)$ receive extra contributions, and  $F_{L}^{D}(3)$
is anomalously large at small $Q^{2}$ because of the nonconservation
of weak currents, for the similar findings in standard DIS see 
\cite{BGNPZcc}.  Second, in the large-$\beta$
CC DDIS with $e^{+}$ beams one excites the $c\bar{s}$ pairs
and the pQCD scale $q_{T}^{2}$ is different for DDIS with leading
charm, $q_{T}^{2} \sim m_{s}^{2}/(1-\beta)$ and leading 
(anti)strangeness, $q_{T}^{2} \sim m_{c}^{2}/(1-\beta)$. Third, the 
$k^{2}$ distribution of the charm changes from the forward to
backward hemispheres in the diffractive system.


\section*{Diffractive factorization, QCD evolution and higher twists}

The interpretation of $q\bar{q}$ excitation as DIS off the intrinsic 
$q\bar{q}$ of the pomeron must be taken with the grain of salt, 
because for the above variety of pQCD scales ($\mu_{G}^{2},m_{f}^{2},
{1\over 4}Q^{2}$), the flavor, valence, sea and glue 
composition of the pomeron changes with $x_{\Pom}$ \cite{GNZ95}. 
Furthermore, because the pQCD scales $q_{T,L}^{2}$
depend on $\beta$, and the $x_{\Pom}$ and $\beta$ 
dependences are inextricably entangled, it is crystal 
clear that the often
postulated \cite{Berera}, but  never proven, diffractive and/or 
Ingelman-Schlein factorization, $F_{2}^{D(3)}(x_{\Pom},\beta,Q^{2})=
f_{\Pom}(x_{\Pom}) F_{2\Pom}(\beta,Q^{2})$, with the process independent 
flux of pomerons in the proton 
$f_{\Pom}(x_{\Pom})$ and the $x_{\Pom}$ independent
structure function $F_{2\Pom}(\beta,Q^{2})$ {\bf does not exist in QCD.}

Besides this diffractive factorization breaking,
we predicted an unprecedented
dominance of the higher twist $F_{L}^{D(3)}$ at 
$\beta \gsim 0.9$, which makes the DGLAP evolution of $F_{2}^{D}$ 
completely invalid at large $\beta$. Still further, upon the 
$k^{2}$ integration, the factor $[1- 2(k^{2}+m_{f}^{2})/M^{2}]$ in 
(\ref{eq:2.1}) transforms into the factor $\sim \left[1-{m_{f}^{2} 
\over Q^{2}(1-\beta)}  G^{2}(x_{\Pom},{1\over 4\beta}Q^{2})\right]$ 
in $F_{T}^{D(3)}$ \cite{BGNZtwist}. Such an abnormal enhancement 
of a higher twist, $\propto G^{2}(x_{\Pom},{1\over 4\beta}Q^{2})$, is 
unprecedented in DIS. This higher twist is especially large 
at $\beta \rightarrow 1$ and is a further strong objection
to the blind DGLAP evolution of $F_{T}^{D(3)}$. This higher twist
is a likely 
explanation \cite{BGNZtwist} of the H1 finding 
\cite{Dirkmann}
of the rise of $F_{2}^{D(3)}$ with $Q^{2}$ even for large $\beta$,
for the related discussion see also \cite{WusthoffDIS97}. 
In the CC DDIS there are still more higher twists
which are due to 
the nonconservation of charged currents \cite{BGNZcc}. 

The non-negotiable prediction from pQCD is a substantial intrinsic
charm in the pomeron \cite{NZ92,GNZ95,GNZcharm}, the abundance of which
for $\beta \gsim 0.1$ is predicted to rise from $\approx $3\% at
$x_{\Pom}=10^{-2}$ to $\approx 25\%$ at $x_{\Pom} = 10 ^{-4}$. 
We predicted the charm abundance
$\sim 15\%$ at $\beta \ll 1$ from excitation
of the $c\bar{c}g$ and higher states. This
intrinsic charm explains perfectly the diffractive charm data reported
at DIS'97 \cite{ZEUScharm,H1charm} without invoking unnatural hard
glue in the pomeron.

The above findings did not make their way yet into the widely used 
Monte Carlo codes for DDIS, which are
all still based on the discredited diffractive factorization and 
the DGLAP evolution of the large-$\beta$
diffractive structure function.
As we have seen above, the transverse momentum structure of 
final states in DDIS is quite different from that in the standard 
DIS. An analysis \cite{NZ94} of higher orders in DDIS has shown
that the structure of real and virtual radiative corrections in
DDIS and standard DIS is different, and one needs more work on
final state radiation before turning on hadron shower codes
which were developed for, and fine tuned to, the standard DIS
final states.
In conclusion, all the DGLAP analysis and the factorization 
model-based conclusions on the hard glue in the pomeron are suspect. 
One badly needs Monte-Carlo implementation of the color dipole 
picture with due incorporation of the above host
of pQCD scales, which is not available at
the moment. 

The situation changes at $\beta\ll 1$, where $F_{2}^{D(3)}(x_{\Pom},
\beta,Q^{2})$ has been proven to satisfy the conventional DGLAP
evolution in $\beta,Q^{2}$ (at least for a fixed $x_{\Pom}$) \cite{NZ94}.
The obvious conclusions are the rise of $F_{2}^{D(3)}(x_{\Pom},\beta,Q^{2})$
as $\beta$ decreases and/or $Q^{2}$ rises and (still approximate)
applicability of the
conventional parton model for hard 
scattering off the pomeron.

 Incidentally, via unitarity  
the diffractive higher twist also implies substantial
higher twist effects in the proton structure function at $x\ll 1$,
which hitherto has been ignored in the QCD analysis of the HERA data.

\section*{Triple-Regge revisited}

The real reason behind the plethora of pQCD scales and effective 
intercepts $\alpha_{\Pom}$ which vary with $Q^{2},\beta,x_{\Pom}$ and
flavor, 
is that the QCD pomeron described by the running BFKL equation
is a sequence of {\bf moving} poles with 
intercepts $\Delta_{n}=\Delta_{\Pom}/(n+1)$ 
\cite{Lipatov,NZZdelta,ZollerDIS97}.
Here $n$ labels the eigenfunctions $\sigma_{n}(r)$ of the running 
BFKL equation for the dipole cross section 
and equals the number of nodes in $\sigma_{n}(r)$, for more 
details see Zoller's talk at DIS'97 \cite{ZollerDIS97}. 
The onset of the dominance of the rightmost pole with $\alpha_{\Pom}
=1+\Delta_{\Pom}$ is elusive because of the substantial contribution
from the subleading poles in the range of $x$ accessible at 
HERA \cite{NZHera,NZZdelta}.
\begin{figure}[t!] 
\centerline{\epsfig{file=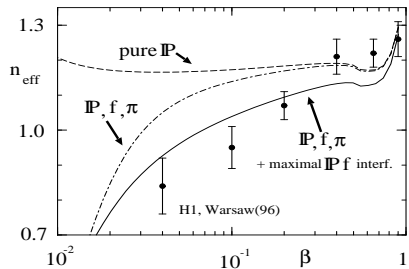,height=1.4in,width=2.1in}}
\vspace{10pt}
\caption{The effective exponent $n_{eff}$ in the fit $F_{2}^{D(3)}
\propto x_{\Pom}^{1-n_{eff}}$ for the pure pomeron contribution to
$F_{2}^{D(3)}$ ({\protect\cite{GNZ95}}, dashed curve) and with 
allowance for the $f$ and $\pi$ exchange in the weak $f\-\Pom$
(dot-dashed curve) and strong $f\-\Pom$ interference scenarios
{\protect\cite{NSZtriple}}.}
\label{fig2}
\end{figure}

Besides the subleading pomeron poles,
DDIS at a moderately small $x_{\Pom}$ picks 
secondary reggeon exchanges $(f,\pi)$. The $f$ exchange
in the $\gamma^{*}p$ scattering originates from DIS off 
the valence quarks and is represented by color singlet
$q\bar{q}$ exchange in the $t$-channel. Because of the kinematical
relation, $x_{\Pom}=x/\beta$, the $f$-reggeon exchange is most 
prominent in DDIS at 
small $\beta$ and lowers the exponent $n_{eff}$ in the fit
$F_{2}^{D} \propto x_{\Pom}^{1-n_{eff}}$, which is evident in the discussion
of the proton spectra in \cite{HERApion} and has indeed been 
observed by the H1 \cite{Dirkmann}. 

Eventually, one must be able to relate the $f$-exchange 
in DDIS to the valence structure function of the proton 
and to calculate the structure function of the reggeon directly from 
pQCD \cite{NSZvalence}. The existing triple-Regge analyses 
\cite{Dirkmann,NSZtriple,Golectriple} rather use the model estimates for 
the flux of reggeons from hadronic diffraction and take for the
reggeon the pion structure function. The major issue is the
reggeon-pomeron interference. If one treats the pomeron 
and reggeon as "orthogonal hadronic
states", then the interference must be weak \cite{NSZtriple}.
To the contrary, the pQCD  suggests \cite{NSZvalence} 
a strong constructive reggeon-pomeron interference. The strong 
interference scenario is preferred by the H1 data shown in Fig.~2.


\section*{Forward diffraction cone in DDIS}

The $p_{\perp}^{2}$ dependence of diffractive cross section is
a new observable accessible with the advent of the  LPS era.
The standard parameterization is 
$d\sigma^{D} \propto \exp(-B_{d}p_{\perp}^{2})$. The so-defined
diffraction slope $B_{d}$ is a fundamental
measure of the interaction radius.
For the two-body scattering  $ac\rightarrow bd$,
an essentially model-independent decomposition is 
$B_{d}=\Delta B_{ab}+\Delta B_{cd} + \Delta B_{int}$, where 
$\Delta B_{ij}$ come from the size of the 
$ij$ transition vertex and $\Delta B_{int}$ comes from the 
interaction range proper. The values of $\Delta B_{ij}$ depend 
strongly on the excitation energy in the $i\rightarrow j$ transition. 
The situation in hadronic diffraction can be summarized 
as follows (\cite{HoltmannDD}, see also the review  
\cite{DDhadronic}): In the elastic 
case, $i=j$, and for excitation of resonances and of the 
low-mass continuum states, 
$\Delta m \lsim m_{N}$, 
which fall into the broad category of {\sl exclusive}
diffraction,
one finds $\Delta B_{ij} \sim \Delta B_{ii}
\approx {1\over 3}R_{i}^{2}$ and $B_{d} \sim B_{el}$,
Here $R_{i}^{2}$ is the hadronic radius squared. In the hadronic
elastic scattering $\Delta B_{ii}\sim$ 4-6 GeV$^{-2}$, and typically
$B_{el}\sim 10$ GeV$^{-2}$. But if $\Delta m \gsim m_{N}$,{\sl i.e.},
then $\Delta B_{ij} \sim 0$. For instance, 
only the target proton size, and $\Delta B_{int}$, contribute 
to $B_{d}$ in the triple-pomeron region of truly {\sl inclusive}
diffraction $hp\rightarrow Xp'$ summed over all large-mass diffractive 
states, $M^{2} \gsim $(5-10)
GeV$^{2}$.  Here one finds  
$B_{d}=B_{3\Pom}=\Delta B_{pp} +\Delta B_{int} \sim {1\over 2}B_{el} 
\approx $6 GeV$^{-2}$. In the double high-mass diffraction
$hp \rightarrow XY$, when $M_{X,Y} \gg m_{N}$, one is left with
$B_{d} \sim \Delta B_{int}\sim $1.5-2 GeV$^{-2}$ (for the especially
illuminating data on double diffraction see \cite{Conta}). The
Regge shrinkage of the diffraction, {\sl i.e.}, the Regge rise of 
$B_{el}$ with energy, was seen in all elastic scattering processes; 
there is as yet no clear evidence for the shrinkage of the diffraction
cone for diffraction dissociation.

The remarkable feature of $B_{3\Pom}$ is its universality,
because the dependence on the diffracting beam and diffractively
produced state drops out; the same $B_{3\Pom}\sim 6$ GeV$^{-2}$
is found for $h=p,\pi,K$ and for real photoproduction $\gamma p
\rightarrow Xp'$ \cite{Chapin}. Furthermore, the absence of the
$Q^{2}$ dependence and the same $B_{d}
\approx B_{3\Pom}$ has been argued to hold for DDIS 
at $M^{2}\gg Q^{2}$, {\sl i.e.}, at $\beta \ll 1$ \cite{NZ92}, 
which agrees with the first data from ZEUS  LPS:
$B_{d} = 7.1\pm 1.1^{+07}_{-1.0}$ GeV$^{-2}$ in DDIS for
5 GeV$^{-2} <Q^{2}< $20 GeV$^{-2}$ 
\cite{Grothe} and $B_{d}=7.7\pm 0.9 \pm 1.0$ GeV$^{-2}$ in real
photoproduction \cite{Briskin}. 
  
DDIS at moderate and large $\beta$ is the tricky one.
Here the principal issues are: i) what is the typical dipole size
(scanning radius) in the $\gamma^{*}X$ transition, and ii) where 
is the borderline between {\sl exclusive} and {\sl inclusive}
DDIS, {\sl i.e.}, what is the relevant excitation scale - the hadronic scale
$\Delta m$ or $\sqrt{Q^{2}}$ ?

\begin{figure}[t!] 
\centerline{\epsfig{file=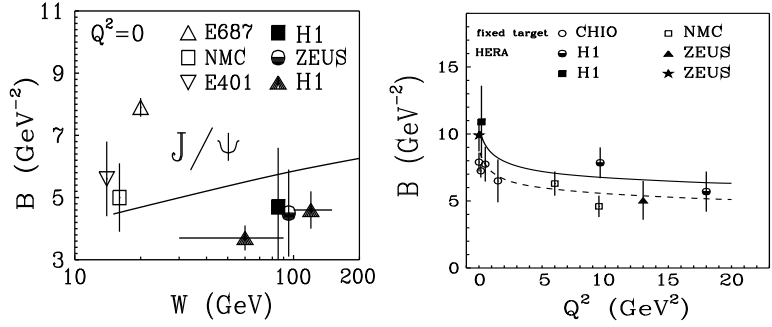,height=1.7in,width=4.in}}
\vspace{10pt}
\caption{Predictions from the color dipole approach for diffraction
slope {\protect\cite{NNPZZslope}} vs. the experimental data.
The left box: the Regge growth of the diffraction slope 
for real photoproduction of the $J/\Psi$. The right box: the $Q^{2}$
dependence of the diffraction slope for the $\rho^{0}$ production.
}
\label{fig3}
\end{figure}

We start with extreme case of exclusive diffraction:  elastic 
production of vector mesons $\gamma^{*}p\rightarrow Vp$. 
The major predictions from the color dipole dynamics for 
elastic production are \cite{NZZslope,NNPZZslope}:
i) The QCD pomeron is a series of {\bf moving} poles, 
the Regge shrinkage persists at all $Q^{2}$ and for all vector mesons 
we
predict the rise of $B_{d}$ 
by $\sim 1.5$\,GeV$^{-2}$ from the CERN/FNAL 
to HERA energy (Fig.~3). ii) Because of the shrinkage of the photon
and the decrease of the scanning radius $r_{S}$ with $Q^{2}$, we
predict $\Delta B_{\gamma^{*}V} \propto r_{S}^{2} \propto 
1/(Q^{2}+m_{V}^{2})$, in agreement with the experiment, 
see Fig.~3 (a more detailed analysis reveals a nontrivial
slowly decreasing contribution to $\Delta B_{\gamma^{*}V}$ which,
however, is numerically small: $\Delta B_{\gamma^{*}V} \sim
R_{c}^{2}/G(x,\tau(Q^{2}+m_{V}^{2})$ \cite{NNPZZslope}). iii) In 
the proton dissociative vector meson production one expects 
\cite{HoltmannDD}
$B_{d} \sim B_{int}$ plus a contribution from the $\gamma^{*}V$
transition vertex which is small at large $Q^{2}$, which has been
confirmed experimentally. For instance, H1 found 
$B_{d}=2.1\pm 0.5\pm 0.5$ for the proton dissociative electroproduction
of the $\rho^{0}$ at $Q^{2} > 7$ GeV$^{-2}$  \cite{H1pdif} and
$B_{D}=1.8\pm 0.3 \pm 0.1$ GeV$^{-2}$  \cite{H1JPsi}. iv) Restoration
of flavor symmetry is predicted, {\sl i.e.,} the values
of $B_{D}$ for different vector mesons must be to a good accuracy 
equal if production of different vector mesons is compared
at equal $Q^{2}+m_{V}^{2}$.

Evidently, in the elastic production $B_{d} \gsim \Delta B_{N} 
\sim $(4-6) GeV$^{-2}$. The HERA results give much too small 
$B_{d}(\gamma^{*} \rightarrow J/\Psi)$. A likely explanation is 
an admixture of proton diffractive $\gamma^{*}p\rightarrow J/\Psi+Y$, 
in which $B_{d} = \Delta B_{int}\sim$1.5-2 GeV$^{-2}$. The LPS data 
can clarify the situation. If an anomalously small values of
$B_{d}(\gamma^{*} \rightarrow J/\Psi)$ will persist, it will 
mean we don't understand the size of protons as probed by 
gluons and will put in trouble many a models of diffractive production. 

\begin{figure}[t!] 
\centerline{\epsfig{file=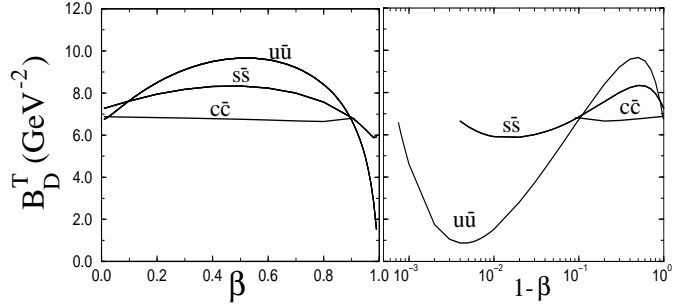,height=1.6in,width=3.5in}}
\vspace{10pt}
\caption{Predictions for the diffraction slope for $\sigma_{T}^{(3)}$
vs. $\beta$  
{\protect\cite{NPZslope}}.}
\label{fig4}
\end{figure}

Because of the node in the radial wave function of radially
excited $V'(2S)$, diffractive production $\gamma^{*}p\rightarrow
V'(2S)p'$ is predicted to have many peculiarities: i) strong
suppression of the $V(1S)/V'(2S)$ production ratio 
\cite{KZ91,KNNZ93,NNPZlight}, which has been well confirmed 
experimentally,
ii) a counterintuitive $B_{d}(\gamma^{*} \rightarrow V'(2S)) 
< B_{d}(\gamma^{*} \rightarrow V(1S))$, 
although $V'(2S)$ has a larger radius, $R(2S) \sim 2 R(1S)$ \cite{NNPZZslope}.
 For instance, for the $\rho'(2S)$
production even the forward dip is possible!
These predictions can be tested soon at HERA.

DDIS at finite $\beta$ is excitation of continuum states,
$M^{2}=Q^{2}(1-\beta)/\beta \gg m_{V}^{2}$. If excitation 
of any continuum were an {\sl inclusive} diffraction, then 
one could have expected 
$B_{d} \approx B_{3\Pom}$, as in the large-mass 
hadronic diffraction and/or large-mass real photoproduction.
This is not the case in DDIS; careful analysis of the $q\bar{q}$ excitation
revealed \cite{NPZslope}  nontrivial 
variation of
$B_{d}^{T}$ about $B_{3\Pom}$, shown in Fig.~4. 
First, the excitation scale is set by $\sqrt{Q^{2}}$ 
rather than the quarkonium mass, so that the diffraction slope 
at fixed $\beta$ must
not depend on $Q^{2}$ . Second, for 
$\beta \rightarrow 0$ one indeed finds an approximately flavor
independent $B_{d}^{T}(\beta \ll 1) \approx B_{3\Pom}$, whereas for
light flavors $B_{d}^{T}(\beta \sim 0.5)$ is large, similar to 
the real photoproduction value of $B(\gamma\rightarrow \rho,\phi)$,
as it has been anticipated
in \cite{NZ94,GNZ95}.
Third, for heavy flavors the $\beta$-dependence is weaker, because the
effective dipole size in the $\gamma^{*}\rightarrow X$ vertex is
smaller: $r^{2} \sim (1-\beta)/ m_{f}^{2}$. Fourth, the decrease 
of $r^{2}$ with $\beta$ explains the decrease of $B_{d}(\beta)$ at
$\beta \rightarrow 1$. 

The predictions \cite{NPZslope} for $\beta 
\rightarrow 1$, shown in Fig.~4,  are
extremely interesting from the point of view of the exclusive-inclusive
duality \cite{GNZlong}. This is a legitimate pQCD region, although a
better
treatment of the final state $q\bar{q}$ interaction is called.
In the extreme limit of $\beta \rightarrow 1$, {\it i.e.,} $M\sim 2m_{f}
\sim m_{V}$,
the relevant dipole size is given by a small scanning radius $r_{S}$
and we predict $B_{d}^{T}\approx B_{d}(\gamma^{*} \rightarrow V)$.
At $1-\beta \sim m_{V}^{2}/Q^{2}$
we predict a substantial drop of the diffraction slope. The drop
is $\propto 1/m_{f}^{2}$ and comes from the rise of the 
quark helicity changing $\gamma^{*}
\rightarrow X$ transition with the momentum transfer.
The dropping $B_{d}$ nicely correlates with the prediction 
\cite{NNPZZslope} of a small diffraction slope for $V'(2S)$ production.
To test this prediction, it is necessary to experimentally
separate the $\sigma_{L}$ from  $\sigma_{T}$ contributions. 
Otherwise the observed cross section is 
dominated  by $\sigma_{L}^{D(3)}$ for which
$B_{d}^{L}\approx B_{3\Pom}$, which is uninteresting. 
These predictions can be 
tested soon with the LPS data from  ZEUS and H1.


\section*{L/T separation in DDIS}

Testing the fundamental pQCD prediction \cite{GNZlong} of the  
dominance of $\sigma_{L}^{D(3)}$ at large $\beta$ is crucial 
for the QCD interpretation of DDIS. The measurement of 
$\sigma_{LT}$ with LPS is a good substitute for the 
measurement of $R=\sigma_{L}/\sigma_{T}$ by varying the beam energies,
which is not possible at the moment. The key is the determination
of the quark helicity changing and conserving amplitudes $\Phi_{1}$ 
and $\Phi_{2}$ \cite{NZ92}. They both contribute
to $d\sigma_{T}^{D(3)}$, whereas $d\sigma_{L}^{D(3)} \sim {1\over Q^{2}}
\Phi_{2}^{2}$
and $d\sigma_{LT}^{D(3)} \propto {p_{\perp} \over Q}\Phi_{1} \Phi_{2}$.
The azimuthal $\cos\phi$ asymmetry is proportional to 
$
A_{LT} = d\sigma_{LT}^{D(3)}/(d\sigma_{T}^{D(3)}+d\sigma_{L}^{D(3)})\, .
$
It provides a stringent test of the pQCD mechanism of DDIS.
In Fig.~5 we show the pQCD Born approximation for $A_{LT}$ 
at $x_{\Pom}\sim 10^{-2}$ and $p_{\perp}^{2}=0.25$ GeV$^{2}$;
the higher orders do not change the gross features of $A_{LT}$
\cite{NPZslope}.
For the $\sigma_{T}$-dominated region of $\beta < $0.85-0.9, the
asymmetry $A_{LT}$ decreases with $Q^{2}$. In contrast, 
it rises with $Q^{2}$
for the $\sigma_{L}$-dominated  $\beta > $0.9. The predicted
asymmetry is quite large and is measurable with LPS.

\begin{figure}[t!] 
\centerline{\epsfig{file=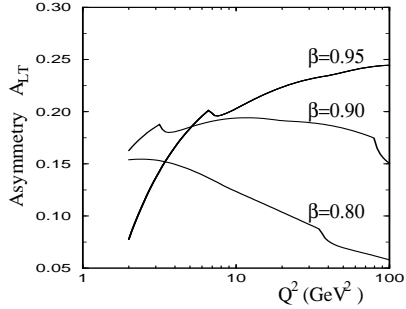,height=1.6in,width=2.1in}}
\vspace{10pt}
\caption{Predictions for the $(e,e')$ plane - recoil proton 
azimuthal asymmetry {\protect\cite{NPZslope}} The spikes on 
curves are due to flavor thresholds at
$Q^{2}=4m_{f}^{2}\beta/(1-\beta)$.
.}
\label{fig5}
\end{figure}

Azimuthal asymmetries
for $q\bar{q}$ dijet production offer more 
tests of the mechanism of DDIS.
One interesting prediction \cite{NPZslope} is a dependence of 
$B_{d}^{T}$ on the  angle $\psi$ between 
the $q\bar{q}$ plane and $\vec{p}_{\perp}$: $B_{d}^{T}(\psi)\approx 
B_{d}(1+\delta_{T}\cos2\psi)$. For light flavors  we predict 
quite a 
substantial asymmetry, $\delta_{T}\sim 0.3$, with 
a nontrivial $\beta$ dependence which is different 
for $\sigma_{L}$ and $\sigma_{T}$. For heavy flavors $\delta_{T}$ 
is smaller. The $\beta$ and 
flavor dependence of $\delta_{T,L}$  can be tested
even without identification of dijets. It is sufficient to study
the azimuthal correlation between the recoil protons and any secondary
hadron in the $\gamma^{*}$ debris.

The azimuthal correlation between the $(e,e')$ plane and the $q\bar{q}$
dijet plane comes from the $d\sigma_{TT'}$ term in (\ref{eq:1.1}). 
The sign of the 
$\cos 2\phi$ term is opposite to that for the photon-gluon 
fusion-dominated jet production in the standard DIS, which is an important
signature of the QCD mechanism of DDIS \cite{Bartelsazim}.

\section*{Conclusions}

The pQCD mechanism of DDIS is well understood. New breakthroughs are
expected with the forthcoming LPS data. Fundamental 
predictions for the diffraction
slope, azimuthal asymmetries and $\sigma_{L}$ can be tested.
Abnormally large higher twist effects, a vast variety of pQCD 
scales, the resulting breaking of diffractive 
factorization and inapplicability of the DGLAP evolution to 
diffractive structure functions have been firmly established 
theoretically. These findings did not make their way yet into 
Monte Carlo codes, which are still based on the discredited 
factorization approximation. Furthermore, final state radiation
in DDIS and standard DIS is different and conclusions from 
Monte Carlo comparisons can be quite misleading. 
The principal remaining task is a
derivation of the real and virtual radiative corrections to DDIS
at moderate and large $\beta$. We hope to hear more on that at
DIS'98 in Brussels.\\

It is a pleasure to thank the organizers of DIS'97 for an exciting 
meeting. NNN is grateful to Danny Krakauer for helpful comments
on the manuscript.

\end{document}